\documentclass[aps,prb,reprint,superscriptaddress,amsmath,amsfonts]{revtex4-1}

\usepackage{xcolor}
\definecolor{darkgreen}{rgb}{0,0.6,0}
\definecolor{cyan}{rgb}{0,0.7,0.8}
\usepackage[colorlinks=true,citecolor=darkgreen,urlcolor=cyan]{hyperref}
\usepackage{graphicx}
\usepackage{amsthm}

\theoremstyle{definition}

\newcommand{\mrm}[1]{\mathrm{#1}}

\newcommand{\fref}[1]{Fig.~\ref{#1}}

\newcommand{\sref}[1]{Sec.~\ref{#1}}

\newcommand{\aref}[1]{Sec.~\ref{#1}}

\newcommand{\psref}[1]{\protect{Sec.~\ref{#1}}}

\newcommand{\rcite}[1]{Ref.~\onlinecite{#1}}

\newcommand{\MATLAB}{\textsc{Matlab}}
\newcommand{\ttt}[1]{\texttt{#1}}

\newcommand{\z}{}

\newcommand{\Q}{Y}

\begin{document}

\title{NCON: A tensor network contractor for MATLAB}

\author{Robert N. C. Pfeifer}
\email{rpfeifer@perimeterinstitute.ca}
\affiliation{Perimeter Institute for Theoretical Physics, 31 Caroline St. N, Waterloo, Ontario N2L 2Y5, Canada}  
\author{Glen Evenbly}
\email{evenbly@caltech.edu}
\affiliation{Institute for Quantum Information and Matter, California Institute of Technology, Pasadena CA 91125, USA}
\author{Sukhwinder Singh}
\email{sukhwinder.singh@mq.edu.au}
\affiliation{Center for Engineered Quantum Systems; Dept. of Physics \& Astronomy, Macquarie University, NSW 2109, Australia}
\author{Guifre Vidal}
\affiliation{Perimeter Institute for Theoretical Physics, 31 Caroline St. N, Waterloo, Ontario N2L 2Y5, Canada}  \date{\today}

\begin{abstract}
A fundamental process in the implementation of any numerical tensor network algorithm is that of \emph{contracting} a tensor network. In this process, a network made up of multiple tensors connected by summed indices is reduced to a single tensor or a number by evaluating the index sums. This article presents a \MATLAB{} function \ttt{ncon()}, or ``Network CONtractor'', which accepts as its input a tensor network and a \emph{contraction sequence} describing how this network may be reduced to a single tensor or number. As its output it returns that single tensor or number. The function \ttt{ncon()} may be obtained by downloading the source of this preprint.
\end{abstract}

\maketitle

\section{Introduction}

Tensor network algorithms are a set of extremely powerful tools for the numerical simulation of quantum many-body systems. 
Examples include White's Density Matrix Renormalization Group (DMRG) algorithm,\cite{white1992} which may be understood as a variational algorithm for optimization of the Matrix Product State (MPS) Ansatz and is arguably the preferred numerical method for the study of one-dimensional quantum systems, and more recent proposals such as Projected Entangled Pair States (PEPS)\cite{verstraete2004,jordan2008} and the Multi-scale Entanglement Renormalization Ansatz (MERA)\cite{vidal2007,cincio2008,evenbly2009b,vidal2010} which offer the potential for scalable simulation of two dimensional quantum systems.
Similarly important are several schemes for the contraction of a 2D tensor network representing either a partition function or a scalar product of two tensor network states, such as the Corner Transfer Matrix method (CTM),\cite{nishino1997} the Tensor Renormalization Group (TRG),\cite{levin2007} and generalizations thereof.\cite{gu2008,gu2009,zhao2010}

A task common to all of these algorithms is the need to evaluate the product of multiple tensors sharing one or more summed indices, frequently referred to as \emph{contracting a tensor network}. Where two tensors share one or more summed indices, the evaluation of the sum over these indices is termed a \emph{pairwise contraction}, and where two tensors are combined which do not share any indices, this is termed a \emph{pairwise outer product}. Any tensor network may be contracted by means of a sequence of pairwise contractions and pairwise outer products, collectively called a \emph{pairwise contraction sequence}, but the computational cost of this process may vary with the sequence chosen. It is known\cite{pfeifer2013a} that pairwise sequences are generally to be preferred over operations involving more than two tensors, and the problem of finding an optimal pairwise contraction sequence is discussed further in \rcite{pfeifer2013a}.

Given both a tensor network and a pairwise contraction sequence, the present article introduces a \MATLAB{} function \ttt{ncon()} which contracts the given tensor network in a manner determined by the specified contraction sequence.

\section{History and motivation\label{sec:history}}

As the problem of contracting a tensor network is so fundamental to the implementation of tensor network algorithms, there have been multiple pieces of software written over the years to perform this task. The function \ttt{ncon()} provided in this preprint %
may be considered a conceptual descendant of G.~Vidal's earlier \ttt{scon()} function, which has enjoyed wide distribution and many informal modifications over the years. One objective in releasing \ttt{ncon()} is to provide a definitive update to \ttt{scon()}, incorporating in a single centralised piece of code all the features which have been added to various branches over time. 

A second reason for releasing \ttt{ncon()} is that many versions of \ttt{scon()} contain code by multiple authors, frequently unattributed and seldom documented. In contrast the file \ttt{ncon.m} which implements the \ttt{ncon()} function is of known provenance, having been written from scratch by R.N.C.~Pfeifer, and is centrally maintained. While \ttt{ncon()} does not make use of any pre-existing code, it is nevertheless appropriate to acknowledge where previous work has influenced the way in which \ttt{ncon()} is implemented:

\begin{itemize}
\item As mentioned above, the idea for \ttt{ncon()} has its origins in the earlier function \ttt{scon()} (``Sequential CONtractor'') by G.~Vidal.
\item The method of efficiently implementing pairwise tensor contractions was inspired by the \ttt{con2t()} function of F.~Verstraete. In this function the contraction of a pair of tensors %
is preceded by using the \ttt{permute()} command to collect together all the contracted and non-contracted indices on each tensor, and the \ttt{reshape()} command to combine these indices. The contraction is then realised as a matrix multiplication, which \MATLAB{} performs using the BLAS library. Further reshaping and permuting transforms the resulting matrix back into a tensor. This approach has become a \emph{de facto} standard for the performance of tensor contractions in \MATLAB{}.
\end{itemize}

Although the syntax of \ttt{ncon()} has deliberately been kept the same as that of \ttt{scon()}, allowing for maximum backward compatibility, with \ttt{ncon()} we have taken the opportunity to release a more fully-featured network contractor
including support for disjoint networks, trivial (dimension 1) indices, 1D (and, under limited circumstances, 0D) objects, traces, outer products, and the zeros-in-{sequence} notation used by \ttt{netcon()} in \rcite{pfeifer2013a}.

\section{Usage}

\subsection{Obtaining the reference implementation\label{sec:obtain}}

\begin{enumerate}
\item While viewing the abstract page for the latest version of this arXiv preprint, click ``Download: Other formats''.
\item Click ``Download source''.
\item Save the resulting file with extension ``.tar''. This file is an archive containing both the reference implementation and the \LaTeX{} source for the instructions you are reading. (Note: The arXiv download page states that the file will be downloaded in .tar.gz format. This is incorrect; it is in .tar format only, and will not require unzipping with gzip.)
\item Unpack the archive using your preferred unarchiver (on a UNIX system you could use \texttt{tar xvf filename.tar}).
\end{enumerate}
The \MATLAB{} function \ttt{ncon()} is provided by the file \texttt{ncon.m}. 

\subsection{Using the reference implementation\label{sec:usage}}

Invocation of the \MATLAB{} function \ttt{ncon()} takes the form
\begin{align*}
\ttt{out~=} &~\ttt{ncon(tensorList,legLinks,sequence,}%
\\&\ttt{~~~~~~~~~~~~~~~~~~~~~~~~~~~~~finalOrder);}
\end{align*}
where the input parameters are specified as follows:

\ttt{tensorList} is a $1\times n$ cell array containing the $n$ tensors which make up the tensor network.

\begin{figure}
\includegraphics[width=246.0pt]{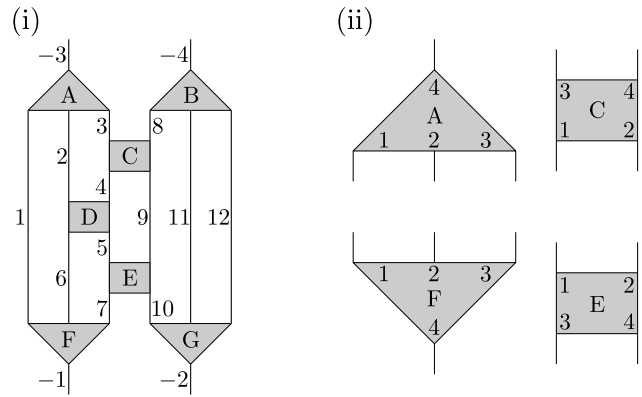}
\caption{(i)~A tensor network diagram arising in the optimization of the 3:1 1D MERA. (ii)~Ordering of indices on the tensors of diagram~(i). Ordering for B is the same as for A. Ordering for D is the same as for C. Ordering for G is the same as for F. Note that ordering for tensor E differs from C and D because tensor E is customarily obtained from tensor C in the 3:1 1D MERA by complex conjugation and vertical reflection, and the process of reflection affects the leg ordering. %
\label{fig:labelledMERA}}
\end{figure}

\ttt{legLinks} describes the tensor network using leg-labelling notation. In brief, the tensor network is represented using the customary diagrammatic notation (for which a summary may be found in \rcite{pfeifer2011b}) and an integer label is assigned to each index (represented in the diagram by a leg). Summed indices are associated with positive integer labels, while open indices are associated with negative integer labels. %
The variable \ttt{legLinks} is then a $1\times n$ cell array with each entry being a row vector whose entries are the integer labels associated with the corresponding tensor. The ordering of these labels matches the ordering of the indices on the corresponding tensor in \MATLAB{}. For example, \fref{fig:labelledMERA}(i) shows a diagram from the 3:1 1D MERA where all indices have been labelled with positive integers. 
A convention is adopted for relating the diagrammatic indices to indices in \MATLAB{}, whereby the indices of a specific \MATLAB{} tensor are associated with specific legs on the diagram. This is illustrated in \fref{fig:labelledMERA}(ii), where (for example) the topmost leg on tensor~A is associated with the fourth index of the \MATLAB{} tensor \ttt{A(:,:,:,:)}. If tensor~A is the first object to appear in \ttt{tensorList} and tensor~B is the second then, reading the labels associated with tensors~A and~B off the diagram of \fref{fig:labelledMERA}(i), \ttt{legLinks} takes the form \ttt{\{[1 2 3 -3],[8 11 12 -4],\ldots\}}.

\ttt{sequence} is a row vector comprising positive integers and (optionally) zero, and specifies a contraction sequence for the closed tensor network. Assuming all indices in \fref{fig:labelledMERA}(i) are of identical dimension, denoted $\chi$, an optimal index contraction sequence for this example network is 
\begin{equation*}
\ttt{[5 4 9 6 7 1 2 3 11 12 8 10]},%
\end{equation*}
having a cost of $2\chi^8+2\chi^7+2\chi^6$. Interpretation of this sequence proceeds as follows: The first entry in the sequence is index~5, connecting tensors D and E. The first step of the contraction sequence is therefore to contract together these two tensors, denoted (D\z{}E). The next entry in the sequence is index~4, connecting the resulting object to tensor C, indicating that the next contraction is  
\begin{equation*}
\mrm{((D\z{}E)\z{}C)}. 
\end{equation*}
This contraction is performed simultaneously over all indices common to both (D\z{}E) and C, and therefore also accounts for index~9.\footnote{Note that when two tensors are contracted together, all indices connecting these tensors should be listed consecutively. Sequences which do not satisfy this condition (i)~are always of suboptimal efficiency (see also \psref{sec:suboptimal}), and (ii)~will result in a warning being generated.}
Proceeding in this fashion for the entirety of the index list, one obtains the pairwise contraction sequence 
\begin{equation*}
\mrm{(((((D\z{}E)\z{}C)\z{}F)\z{}A)\z{}(B\z{}G))}. 
\end{equation*}
Note that if the \ttt{netcon()} reference implementation given in \rcite{pfeifer2013a} is installed then this may be used to automatically 
generate an optimal index-based contraction sequence for a tensor network. %

The \ttt{ncon()} function includes support for contraction sequences involving outer products, either where two tensors are contracted despite not sharing an index or where the tensors share only indices of dimension~1, and the appropriate syntaxes for \ttt{sequence} in these situations are discussed in \aref{sec:outerprod}. 
Further discussion of the index-labelling notation for contraction sequences may also be found in \rcite{pfeifer2013a},
including an explicit
algorithm for converting sequences of index labels into sequences of pairwise tensor contractions.

Note that the argument \ttt{sequence} is optional, and if not specified it will default to the list of all positive indices in ascending numerical order. 

Finally, a fourth argument, \ttt{finalOrder}, which is also optional, may be used to modify the ordering of indices on the output tensor. By default these indices correspond to the negative entries in \ttt{legLinks} in descending numerical order, so that indices 1, 2, 3, 4, etc. of the output tensor correspond to labels \ttt{-1}, \ttt{-2}, \ttt{-3}, \ttt{-4}, etc. respectively. This behaviour may be changed by specifying a sequence for the negative labels in \ttt{finalOrder}, e.g.
\begin{equation*}
\ttt{finalOrder = [-3 -1 -2 -4 \ldots]}. 
\end{equation*}
The first index of the output tensor would then correspond to the index in \ttt{legLinks} which bears label \ttt{-3}, and so forth. As the order of the negatively-labelled legs is explicitly specified, the requirement that these legs be numbered consecutively from $-1$ is also lifted when \ttt{finalOrder} is given.

The \ttt{ncon()} function incorporates substantial error checking of user-supplied input. The computational cost of this is small, but when invoking \ttt{ncon()} from stable code it may be desireable to disable the checks in exchange for a slight increase in performance. This may be achieved by declaring a global variable \ttt{ncon\_skipCheckInputs} and setting its value to true:
\begin{align*}
&\ttt{global ncon\_skipCheckInputs;}\\
&\ttt{ncon\_skipCheckInputs = true;}
\end{align*}
This setting will persist until the variable is cleared from the global workspace or its value is changed to anything other than \ttt{true}. For users unfamiliar with global variables, the above command should be understood as turning off error checking in \ttt{ncon()} until you exit \MATLAB{}, issue a \ttt{clear all} command, or type
\begin{align*}
&\ttt{global ncon\_skipCheckInputs;}\\
&\ttt{ncon\_skipCheckInputs = false;}
\end{align*}
to turn it back on again. To avoid confusion between global and local variables, it is recommended that the variable name \ttt{ncon\_skipCheckInputs} should not be used for any other purpose.

\section{Simple examples}

In this Section, the syntax of \ttt{ncon()} described in \sref{sec:usage} is illustrated with two relatively simple examples.

\subsection{Matrix multiplication}

The first example is that of matrix multiplication, $$C=AB,$$ where $A$ and $B$ are matrices. Using index notation, this calculation may be written
\begin{equation}
\mrm{C}_{\alpha\beta} = \sum_\gamma \mrm{A}_{\alpha\gamma} \mrm{B}_{\gamma\beta}.
\end{equation}
Using \ttt{ncon()}, matrix~C may be constructed using the commands
\begin{align*}
&\ttt{tensors = \{A,B\};}\\
&\ttt{legLinks = \{[-1 1],[1 -2]\};}\\
&\ttt{sequence = 1;}\\
&\ttt{C = ncon(tensors,legLinks,sequence);}
\end{align*}
or more directly,
\begin{align*}
\ttt{C = ncon(\{A,B\},\{[-1 1],[1 -2]\},1);}
\end{align*}

For example, if~A and~B are created using the commands
\begin{align*}
&\ttt{A = rand(3,4);}\\
&\ttt{B = rand(4,6);}
\end{align*}
then the output of the above command is seen to be exactly equal to that obtained by typing 
\begin{equation*}
\ttt{C = A*B;}. 
\end{equation*}
There is seldom any reason to perform a simple matrix multiplication using \ttt{ncon()} instead of the multiplication operator \ttt{*}, but this serves as a simple illustrative example of the syntax of the \ttt{ncon()} command.

\subsection{Matrix and two vectors}

A slightly more complicated example is given by the expression
\begin{equation}
c = x^\mrm{T}My
\end{equation}
where $M$ is a matrix and $x$ and $y$ are column vectors. In index notation,
\begin{equation}
c = \sum_{\alpha,\beta} x_\alpha M_{\alpha\beta} y_\beta.
\end{equation}
Once again, $c$ may be calculated in \MATLAB{} using the multiplication operator
\begin{equation*}
\ttt{c = (x.')*M*y;}
\end{equation*}
or using \ttt{ncon()}. This time the syntax for the \ttt{ncon()} command is
\begin{align*}
&\ttt{tensors = \{x,M,Y\};}\\
&\ttt{legLinks = \{[1],[1 2],[2]\};}\\
&\ttt{sequence = [1 2];}\\
&\ttt{c = ncon(tensors,legLinks,sequence);}
\end{align*}
or more briefly,
\begin{equation*}
\ttt{c = ncon(\{x,M,y\},\{[1],[1 2],[2]\},[1 2]);}
\end{equation*}
where the sequence $[1~2]$ indicates this is to be evaluated as
\begin{equation}
c = (x^\mrm{T}M)y
\end{equation}
in contrast to a contraction sequence of $[2~1]$ which would correspond to
\begin{equation}
c = x^\mrm{T}(My).
\end{equation}
Which sequence is to be preferred will depend upon the relative lengths of vectors $x$ and $y$, with the calculation proceeding more rapidly if the longer vector is multiplied with $M$ first, though the value of $c$ obtained will of course be unaffected by this choice.

\section{Sequences involving outer products\label{sec:outerprod}}

On occasion, the optimal contraction sequence for a tensor network may necessarily involve an outer product of two or more tensors. The \ttt{ncon()} function supports three distinct methods of specifying an outer product. Each has its own benefits and drawbacks.

Note that the discussion of outer products in this Section closely parallels (and partially reproduces) that of Appendix~B in \rcite{evenbly2013}. Readers familiar with the function \ttt{multienv()} presented in \rcite{evenbly2013} are encouraged to read \sref{sec:disjOP} on disjoint networks, as this feature is substantially more powerful in \ttt{ncon()} than in \ttt{multienv()}, and also to review the example invocations of \ttt{ncon()} presented in the rest of the Section and to compare these with their counterparts in Appendix~B of \rcite{evenbly2013}. Readers not familiar with \ttt{multienv()} may be reassured that familiarity with \ttt{multienv()} is not required in order to make effective use of \ttt{ncon()}.

\subsection{Disjoint networks\label{sec:disjOP}}

As the first method of describing a contraction sequence involving outer products, one may specify a disjoint tensor network, e.g.
\begin{align*}
&\ttt{A = rand(2,2);B = rand(2,2);C = rand(2,2);}\\
&\ttt{D = ncon(\{A,B,C\},\{[-1 1],[1 -3],[-2 -4]\},[1])}.
\end{align*}
After contracting over index~1, the resulting network is made up of two disconnected tensors. Writing
\begin{equation}
E_{\alpha\beta}=\sum_\gamma A_{\alpha\gamma}B_{\gamma\beta}
\end{equation}
where $\gamma$ corresponds to index 1, the desired output $D_{\alpha\beta\gamma\delta}$ is given by the outer product
\begin{equation}
D_{\alpha\beta\gamma\delta} = E_{\alpha\gamma}C_{\beta\delta}.
\end{equation}
This situation is recognised by %
\ttt{ncon()}, and the outer product is automatically performed in order to explicitly return the requested object, in this instance $D_{\alpha\beta\gamma\delta}$. This method of specifying an outer product is simple and can even be used when the contraction sequence is empty, e.g.
\begin{equation*}
\ttt{F = ncon(\{A,B,C\},\{[-1 -3],[-2 -4],[-5 -6]\})},
\end{equation*}
corresponding to $F_{\alpha\beta\gamma\delta\epsilon\zeta}=A_{\alpha\gamma}B_{\beta\delta}C_{\epsilon\zeta}$.
Where such an outer product is performed across three or more tensors, as here, it is automatically evaluated in the most efficient manner possible (being a series of pairwise contractions between the tensors of smallest total dimension). The primary limitation of this method is that it is only capable of specifying outer products occuring as the final steps of the contraction sequence.

\subsection{Trivial indices\label{sec:dim1}}

\begin{figure}
\includegraphics[width=246.0pt]{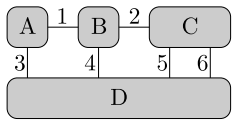}
\caption{Example tensor network with trivial index. All indices have dimension~3 except for the index labelled 2, which has dimension~1.\label{fig:trivindexample}}
\end{figure}

As a second method one may introduce explicit connecting indices of dimension~1. Consider an example, adapted from Appendix~B~2 of \rcite{evenbly2013}, where the dimensions of tensors~B and~C are explicitly $3\times 1\times 3$ and $3\times3\times1$ respectively:\footnote{Note that \MATLAB{} leaves trailing indices of dimension~1 implicit, and so will report the size of tensor~C as $3\times 3$ if queried, and not $3\times 3\times 1$. Equally, it is also valid to create tensor~C using the command \ttt{C~=~rand(3,3);} instead of \ttt{C~=~rand(3,3,1);} as given above.} %
\begin{align*}
&\ttt{A = rand(3,3); B = rand(3,1,3);}\\
&\ttt{C = rand(3,3,1); D = rand(3,3,3,3);}\\
&\ttt{tensors = \{A,B,C,D\};}\\
&\ttt{legs = \{[3 1],[1 2 4],[5 6 2],[3 4 5 6]\};}\\
&\ttt{seq = [1 2 3 4 5 6];}\\
&\ttt{E = ncon(tensors,legs,seq);}
\end{align*}
This tensor network is illustrated in \fref{fig:trivindexample}, 
and the given index contraction sequence corresponds to a pairwise tensor contraction sequence of (((A\z{}B)\z{}C)\z{}D). 
The outer product %
is between %
the contraction product (A\z{}B) and the tensor~C, and corresponds to contraction over the index of dimension~1 carrying the numerical %
label \ttt{2}. This label appears in the index-based contraction sequence in the usual manner.

The inclusion of trivial indices %
is the most versatile means of specifying an outer product %
as part of the contraction sequence%
, with the only drawback being the need to explicitly include the indices of dimension~1 over which the contractions are to be performed.

\subsection{Zeros-in-\ttt{sequence} notation\label{sec:zis}}

\begin{figure}
\includegraphics[width=246.0pt]{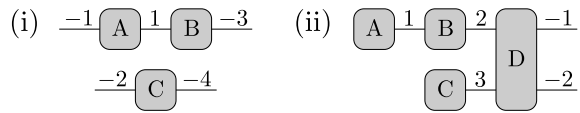}
\caption{Example tensor networks; all indices have dimension~2. (i)~An index sequence of \ttt{[1 0]} corresponds to the pairwise tensor contraction sequence ((A\z{}B)\z{}C) where the second contraction to be performed is an outer product. (ii)~An index sequence of \ttt{[1 0 2 3]} corresponds to the pairwise tensor contraction sequence ((A\z{}B)\z{}C)\z{}D) where, once again, the second contraction to be performed is an outer product.\label{fig:zeroex}}
\end{figure}

In %
\rcite{pfeifer2013a} it was shown that for any tensor network, if an outer product of two or more tensors is \emph{required} as part of the optimal contraction sequence and the result of this outer product is denoted \Q{}, then an optimal contraction sequence may always be found where this outer product is always either
\begin{enumerate}
\item the final step in the contraction of the tensor network, or
\item followed by contracting \emph{all} indices of object~\Q{} with another tensor, which we will denote~X (as per the final Appendix of \rcite{pfeifer2013a}).
\end{enumerate}
Outer products of these forms (and these forms only) may be represented %
by inserting zeros into the contraction sequence, with the outer product of $n$ tensors being indicated by $n-1$ consecutive zeros. To determine the $n$ tensors involved in the outer product when there are more than $n$ tensors remaining, indices are read from the sequence after the zeros, and the tensors carrying these indices are noted, until $n+1$ tensors have been identified. Given the above constraints on the outer products which may be represented using this notation, it then follows that %
one of these $n+1$ tensors will necessarily share summed indices with all $n$ other tensors. This tensor does not participate in the outer product, but is instead the tensor~X which is subsequently contracted with object~\Q{}. For a fuller discussion of this outer product notation, and of the optimal performance of the outer products of multiple tensors, see \rcite{pfeifer2013a}. %
A simple example is given by
\begin{align*}
&\ttt{A = rand(2,2);B = rand(2,2);C = rand(2,2);}\\
&\ttt{D = ncon(\{A,B,C\},\{[-1 1],[1 -3],[-2 -4]\},[1 0]);}
\end{align*}
and illustrated in \fref{fig:zeroex}(i), and a more complicated example is given by
\begin{align*}
&\ttt{A = rand(2,1);B = rand(2,2);C = rand(2,1);}\\
&\ttt{D = rand(2,2,2,2);}\\
&\ttt{tensors = \{A,B,C,D\};}\\
&\ttt{legs = \{[1],[1 2],[3],[2 3 -1 -2]\};}\\
&\ttt{D = ncon(tensors,legs,[1 0 2 3]);}
\end{align*}
where the index sequence \ttt{[1 0 2 3]} corresponds to a tensor contraction sequence (((A\z{}B)\z{}C)\z{}D) and the network is illustrated in \fref{fig:zeroex}(ii). %

While this approach is not as versatile as the use of trivial indices discussed in \aref{sec:dim1}, it is nevertheless useful when an optimal contraction sequence is being automatically generated, for instance by using the Netcon algorithm.\cite{pfeifer2013a} Because there always exists an optimal contraction sequence which may be described using the zeros-in-\ttt{sequence} notation, a search algorithm for optimal sequences can return a valid response in this notation even when all optimal contraction sequences %
involve the performance of
an outer product for which no appropriate 
index of dimension~1 has been provided.

As \ttt{ncon()} supports the zeros-in-\ttt{sequence} notation for outer products, it is capable of directly accepting contraction sequences generated by the reference implementation of Netcon included with \rcite{pfeifer2013a}. Note that when using the zeros-in-\ttt{sequence} notation, all outer products must be explicitly represented in the sequence and not left implicit as in \sref{sec:disjOP}.

\section{How it works}

\subsection{Contraction of tensor networks}

The operation of \ttt{ncon()} is as follows:
\begin{itemize}
\item Basic validity checks are performed on the input data.
\item The main loop of \ttt{ncon()} (which is contained within the function \ttt{performContraction()} in \ttt{ncon.m}) steps through the contraction sequence as follows:
\begin{itemize}
\item If the first entry in the sequence is an index appearing on two tensors, then these tensors are to be contracted.
\begin{itemize}
\item If reading further consecutive entries from the sequence yields additional indices which connect the same two tensors, these indices are also contracted over at the same time (for example indices~4 and~9 in \sref{sec:usage}).
\item If there are other indices connecting these two tensors which appear later in the sequence, then these indices are not contracted over at this time. Such a sequence is suboptimal and a warning is generated. %
\item Pairwise tensor contraction is performed by the function \ttt{tcontract()} in \ttt{ncon.m}.
\end{itemize}
\item If the first entry is an index appearing twice on one tensor, then it is a trace to be performed on that tensor.
\begin{itemize}
\item If reading further consecutive entries from the sequence yields additional indices corresponding to traces on the same tensor, these indices are also traced over at the same time.
\item If there are other indices corresponding to traces on the same tensor that appear later in the sequence, then these indices are not contracted over at this time. Such a sequence is suboptimal and a warning is generated. %
\item Evaluation of traces is performed by the function \ttt{doTrace()} in \ttt{ncon.m}.
\end{itemize}
\item If the first entry is a zero, the corresponding outer product is determined according to the algorithm given in Appendix~E of \rcite{pfeifer2013a}.
\begin{itemize}
\item Parsing of zeros-in-\ttt{sequence} notation is performed by the function \ttt{zisOuterProduct()} in \ttt{ncon.m}.
\end{itemize}
\item Whatever the nature of the entry, the indices contracted over (or the zeros which have been processed) are then deleted from the sequence.
\end{itemize}
\item Finally, the indices of the output object are arranged as specified by the negative indices appearing in \ttt{legLinks} and (optionally) the input parameter \ttt{finalOrder}.
\end{itemize}

\subsection{Contraction of tensor pairs\label{sec:suboptimal}}

As noted in \sref{sec:history}, the implementation of pairwise tensor contraction in \ttt{ncon()} is achieved through use of the \MATLAB{} \ttt{permute()} and \ttt{reshape()} commands, and matrix multiplication. %
One consequence of this approach is that objects thus constructed are always themselves full tensors. Consequently, if given an expression such as
\begin{equation}
A^{\alpha\beta\gamma\delta}_\varepsilon B^\zeta_{\beta\gamma\delta\eta}
\end{equation}
one may efficiently contract tensors $A$ and $B$ over indices $\beta$, $\gamma$, and $\delta$ to yield
\begin{equation}
T^{\alpha\zeta}_{\varepsilon\eta} = A^{\alpha\beta\gamma\delta}_\varepsilon B^\zeta_{\beta\gamma\delta\eta},
\end{equation}
but contraction over only a subset of these indices, say $\beta$ and $\gamma$, is intrinsically inefficient, corresponding to computation of all entries in the object
\begin{equation}
T'{}^{\alpha\delta\zeta}_{\varepsilon\theta\eta} = A^{\alpha\beta\gamma\delta}_\varepsilon B^\zeta_{\beta\gamma\theta\eta}
\end{equation}
when we will subsequently be performing the trace
\begin{equation}
T^{\alpha\zeta}_{\varepsilon\eta} = \sum_\delta T'{}^{\alpha\delta\zeta}_{\varepsilon\delta\eta}
\end{equation}
and so only the entries
\begin{equation}
\left.T'{}^{\alpha\delta\zeta}_{\varepsilon\theta\eta}\right|_{\delta=\theta}
\end{equation}
are actually required.
Calculation of the full tensor $T'$ is wasteful of computational resources, and consequently index sequences which do not contract over all shared indices at once are %
suboptimal in both time and memory usage.

It is interesting to compare this approach with the use of \ttt{for} loops to iterate over the entries in $A$ and $B$ as in \rcite{lam1997}. With explicit control over each individual index, rather than computing the entirety of tensor $T'$ one may recognise that a trace will subsequently be performed over indices $\delta$ and $\theta$, and only evaluate the elements 
\begin{equation}
\left.T'{}^{\alpha\delta\zeta}_{\varepsilon\theta\eta}\right|_{\theta=\delta} = \left.A^{\alpha\beta\gamma\delta}_\varepsilon B^\zeta_{\beta\gamma\theta\eta}\right|_{\theta=\delta}. 
\end{equation}
Taking this approach, the unnecessary computational cost associated with evaluating all entries in $T'$ is eliminated. Such an approach is still inefficient in memory, however, as prior to performing the sum it stores a separate entry for each value of $\delta$,
and consequently when contracting a single tensor network it is always preferable to contract over all shared indices simultaneously %
unless the calculation of intermediate objects such as $T'$ is specifically desired.
In \rcite{lam1997} this is addressed by observing that whenever an index is repeated on a single tensor, it is always appropriate to immediately evaluate the sum over that index.

Note that while \ttt{ncon()} is able to detect inefficiencies of this sort and thus may sometimes recognise
if a contraction sequence is suboptimal, the task of identifying an optimal contraction sequence for a given tensor network is non-trivial.\cite{pfeifer2013a} The %
absence of a warning does not indicate that the contraction sequence supplied to \ttt{ncon()} is optimal; merely that no obvious indicators of a suboptimal sequence were detected.

\begin{acknowledgments}
R.N.C.P. thanks the Ontario Ministry of Research and Innovation Early Researcher Awards for financial support. G.E. is supported by the Sherman Fairchild foundation. S.S. acknowledges financial support from the MQNS grant by Macquarie University Grant No. 9201200274. Research at Perimeter Institute is supported by the Government of Canada through Industry Canada and by the Province of Ontario through the Ministry of Research and Innovation. This research was supported in part by the ARC Centre of Excellence in Engineered Quantum Systems (EQuS), Project No. CE110001013.
\end{acknowledgments}

\bibliography{ncon}

\begin{thebibliography}{18}%
\makeatletter
\providecommand \@ifxundefined [1]{%
 \@ifx{#1\undefined}
}%
\providecommand \@ifnum [1]{%
 \ifnum #1\expandafter \@firstoftwo
 \else \expandafter \@secondoftwo
 \fi
}%
\providecommand \@ifx [1]{%
 \ifx #1\expandafter \@firstoftwo
 \else \expandafter \@secondoftwo
 \fi
}%
\providecommand \natexlab [1]{#1}%
\providecommand \enquote  [1]{``#1''}%
\providecommand \bibnamefont  [1]{#1}%
\providecommand \bibfnamefont [1]{#1}%
\providecommand \citenamefont [1]{#1}%
\providecommand \href@noop [0]{\@secondoftwo}%
\providecommand \href [0]{\begingroup \@sanitize@url \@href}%
\providecommand \@href[1]{\@@startlink{#1}\@@href}%
\providecommand \@@href[1]{\endgroup#1\@@endlink}%
\providecommand \@sanitize@url [0]{\catcode `\\12\catcode `\$12\catcode
  `\&12\catcode `\#12\catcode `\^12\catcode `\_12\catcode `\%12\relax}%
\providecommand \@@startlink[1]{}%
\providecommand \@@endlink[0]{}%
\providecommand \url  [0]{\begingroup\@sanitize@url \@url }%
\providecommand \@url [1]{\endgroup\@href {#1}{\urlprefix }}%
\providecommand \urlprefix  [0]{URL }%
\providecommand \Eprint [0]{\href }%
\providecommand \doibase [0]{http://dx.doi.org/}%
\providecommand \selectlanguage [0]{\@gobble}%
\providecommand \bibinfo  [0]{\@secondoftwo}%
\providecommand \bibfield  [0]{\@secondoftwo}%
\providecommand \translation [1]{[#1]}%
\providecommand \BibitemOpen [0]{}%
\providecommand \bibitemStop [0]{}%
\providecommand \bibitemNoStop [0]{.\EOS\space}%
\providecommand \EOS [0]{\spacefactor3000\relax}%
\providecommand \BibitemShut  [1]{\csname bibitem#1\endcsname}%
\let\auto@bib@innerbib\@empty
\bibitem [{\citenamefont {White}(1992)}]{white1992}%
  \BibitemOpen
  \bibfield  {author} {\bibinfo {author} {\bibfnamefont {S.~R.}\ \bibnamefont
  {White}},\ }\href {\doibase 10.1103/PhysRevLett.69.2863} {\bibfield
  {journal} {\bibinfo  {journal} {Phys. Rev. Lett.}\ }\textbf {\bibinfo
  {volume} {69}},\ \bibinfo {pages} {2863} (\bibinfo {year}
  {1992})}\BibitemShut {NoStop}%
\bibitem [{\citenamefont {Verstraete}\ and\ \citenamefont
  {Cirac}()}]{verstraete2004}%
  \BibitemOpen
  \bibfield  {author} {\bibinfo {author} {\bibfnamefont {F.}~\bibnamefont
  {Verstraete}}\ and\ \bibinfo {author} {\bibfnamefont {J.~I.}\ \bibnamefont
  {Cirac}},\ }\href {http://arxiv.org/abs/cond-mat/0407066v1} {}\Eprint
  {http://arxiv.org/abs/cond-mat/0407066v1}
  {arXiv:cond-mat/0407066v1 (2004)} \BibitemShut {NoStop}%
\bibitem [{\citenamefont {Jordan}\ \emph {et~al.}(2008)\citenamefont {Jordan},
  \citenamefont {Or\'us}, \citenamefont {Vidal}, \citenamefont {Verstraete},\
  and\ \citenamefont {Cirac}}]{jordan2008}%
  \BibitemOpen
  \bibfield  {author} {\bibinfo {author} {\bibfnamefont {J.}~\bibnamefont
  {Jordan}}, \bibinfo {author} {\bibfnamefont {R.}~\bibnamefont {Or\'us}},
  \bibinfo {author} {\bibfnamefont {G.}~\bibnamefont {Vidal}}, \bibinfo
  {author} {\bibfnamefont {F.}~\bibnamefont {Verstraete}}, \ and\ \bibinfo
  {author} {\bibfnamefont {J.~I.}\ \bibnamefont {Cirac}},\ }\href {\doibase
  10.1103/PhysRevLett.101.250602} {\bibfield  {journal} {\bibinfo  {journal}
  {Phys. Rev. Lett.}\ }\textbf {\bibinfo {volume} {101}},\ \bibinfo {pages}
  {250602} (\bibinfo {year} {2008})}\BibitemShut {NoStop}%
\bibitem [{\citenamefont {Vidal}(2007)}]{vidal2007}%
  \BibitemOpen
  \bibfield  {author} {\bibinfo {author} {\bibfnamefont {G.}~\bibnamefont
  {Vidal}},\ }\href {\doibase 10.1103/PhysRevLett.99.220405} {\bibfield
  {journal} {\bibinfo  {journal} {Phys. Rev. Lett.}\ }\textbf {\bibinfo
  {volume} {99}},\ \bibinfo {eid} {220405} (\bibinfo {year}
  {2007})}\BibitemShut {NoStop}%
\bibitem [{\citenamefont {Cincio}\ \emph {et~al.}(2008)\citenamefont {Cincio},
  \citenamefont {Dziarmaga},\ and\ \citenamefont {Rams}}]{cincio2008}%
  \BibitemOpen
  \bibfield  {author} {\bibinfo {author} {\bibfnamefont {L.}~\bibnamefont
  {Cincio}}, \bibinfo {author} {\bibfnamefont {J.}~\bibnamefont {Dziarmaga}}, \
  and\ \bibinfo {author} {\bibfnamefont {M.~M.}\ \bibnamefont {Rams}},\ }\href
  {\doibase 10.1103/PhysRevLett.100.240603} {\bibfield  {journal} {\bibinfo
  {journal} {Phys. Rev. Lett.}\ }\textbf {\bibinfo {volume} {100}},\ \bibinfo
  {pages} {240603} (\bibinfo {year} {2008})}\BibitemShut {NoStop}%
\bibitem [{\citenamefont {Evenbly}\ and\ \citenamefont
  {Vidal}(2009)}]{evenbly2009b}%
  \BibitemOpen
  \bibfield  {author} {\bibinfo {author} {\bibfnamefont {G.}~\bibnamefont
  {Evenbly}}\ and\ \bibinfo {author} {\bibfnamefont {G.}~\bibnamefont
  {Vidal}},\ }\href {\doibase 10.1103/PhysRevLett.102.180406} {\bibfield
  {journal} {\bibinfo  {journal} {Phys. Rev. Lett.}\ }\textbf {\bibinfo
  {volume} {102}},\ \bibinfo {pages} {180406} (\bibinfo {year}
  {2009})}\BibitemShut {NoStop}%
\bibitem [{\citenamefont {Vidal}(2010)}]{vidal2010}%
  \BibitemOpen
  \bibfield  {author} {\bibinfo {author} {\bibfnamefont {G.}~\bibnamefont
  {Vidal}},\ }in\ \href {http://arxiv.org/abs/0912.1651v2} {\emph {\bibinfo
  {booktitle} {Understanding Quantum Phase Transitions}}},\ \bibinfo {editor}
  {edited by\ \bibinfo {editor} {\bibfnamefont {L.~D.}\ \bibnamefont {Carr}}}\
  (\bibinfo  {publisher} {Taylor \& Francis},\ \bibinfo {address} {Boca
  Raton},\ \bibinfo {year} {2010})\BibitemShut {NoStop}%
\bibitem [{\citenamefont {Nishino}\ and\ \citenamefont
  {Okunishi}(1997)}]{nishino1997}%
  \BibitemOpen
  \bibfield  {author} {\bibinfo {author} {\bibfnamefont {T.}~\bibnamefont
  {Nishino}}\ and\ \bibinfo {author} {\bibfnamefont {K.}~\bibnamefont
  {Okunishi}},\ }\href {\doibase 10.1143/JPSJ.66.3040} {\bibfield  {journal}
  {\bibinfo  {journal} {J. Phys. Soc. Jpn.}\ }\textbf {\bibinfo {volume}
  {66}},\ \bibinfo {pages} {3040} (\bibinfo {year} {1997})}\BibitemShut
  {NoStop}%
\bibitem [{\citenamefont {Levin}\ and\ \citenamefont {Nave}(2007)}]{levin2007}%
  \BibitemOpen
  \bibfield  {author} {\bibinfo {author} {\bibfnamefont {M.}~\bibnamefont
  {Levin}}\ and\ \bibinfo {author} {\bibfnamefont {C.~P.}\ \bibnamefont
  {Nave}},\ }\href {\doibase 10.1103/PhysRevLett.99.120601} {\bibfield
  {journal} {\bibinfo  {journal} {Phys. Rev. Lett.}\ }\textbf {\bibinfo
  {volume} {99}},\ \bibinfo {pages} {120601} (\bibinfo {year}
  {2007})}\BibitemShut {NoStop}%
\bibitem [{\citenamefont {Gu}\ \emph {et~al.}(2008)\citenamefont {Gu},
  \citenamefont {Levin},\ and\ \citenamefont {Wen}}]{gu2008}%
  \BibitemOpen
  \bibfield  {author} {\bibinfo {author} {\bibfnamefont {Z.-C.}\ \bibnamefont
  {Gu}}, \bibinfo {author} {\bibfnamefont {M.}~\bibnamefont {Levin}}, \ and\
  \bibinfo {author} {\bibfnamefont {X.-G.}\ \bibnamefont {Wen}},\ }\href
  {\doibase 10.1103/PhysRevB.78.205116} {\bibfield  {journal} {\bibinfo
  {journal} {Phys. Rev. B}\ }\textbf {\bibinfo {volume} {78}},\ \bibinfo
  {pages} {205116} (\bibinfo {year} {2008})}\BibitemShut {NoStop}%
\bibitem [{\citenamefont {Gu}\ and\ \citenamefont {Wen}(2009)}]{gu2009}%
  \BibitemOpen
  \bibfield  {author} {\bibinfo {author} {\bibfnamefont {Z.-C.}\ \bibnamefont
  {Gu}}\ and\ \bibinfo {author} {\bibfnamefont {X.-G.}\ \bibnamefont {Wen}},\
  }\href {\doibase 10.1103/PhysRevB.80.155131} {\bibfield  {journal} {\bibinfo
  {journal} {Phys. Rev. B}\ }\textbf {\bibinfo {volume} {80}},\ \bibinfo
  {pages} {155131} (\bibinfo {year} {2009})}\BibitemShut {NoStop}%
\bibitem [{\citenamefont {Zhao}\ \emph {et~al.}(2010)\citenamefont {Zhao},
  \citenamefont {Xie}, \citenamefont {Chen}, \citenamefont {Wei}, \citenamefont
  {Cai},\ and\ \citenamefont {Xiang}}]{zhao2010}%
  \BibitemOpen
  \bibfield  {author} {\bibinfo {author} {\bibfnamefont {H.~H.}\ \bibnamefont
  {Zhao}}, \bibinfo {author} {\bibfnamefont {Z.~Y.}\ \bibnamefont {Xie}},
  \bibinfo {author} {\bibfnamefont {Q.~N.}\ \bibnamefont {Chen}}, \bibinfo
  {author} {\bibfnamefont {Z.~C.}\ \bibnamefont {Wei}}, \bibinfo {author}
  {\bibfnamefont {J.~W.}\ \bibnamefont {Cai}}, \ and\ \bibinfo {author}
  {\bibfnamefont {T.}~\bibnamefont {Xiang}},\ }\href {\doibase
  10.1103/PhysRevB.81.174411} {\bibfield  {journal} {\bibinfo  {journal} {Phys.
  Rev. B}\ }\textbf {\bibinfo {volume} {81}},\ \bibinfo {pages} {174411}
  (\bibinfo {year} {2010})}\BibitemShut {NoStop}%
\bibitem [{\citenamefont {Pfeifer}()}]{pfeifer2013a}%
  \BibitemOpen
  \bibfield  {author} {\bibinfo {author} {\bibfnamefont {R.~N.~C.}\
  \bibnamefont {Pfeifer}},\ }\href {http://arxiv.org/abs/1304.6112} {}\Eprint
  {http://arxiv.org/abs/1304.6112}
  {arXiv:1304.6112 [cond-mat.str-el] (2013)} \BibitemShut {NoStop}%
\bibitem [{\citenamefont {Pfeifer}(2011)}]{pfeifer2011b}%
  \BibitemOpen
  \bibfield  {author} {\bibinfo {author} {\bibfnamefont {R.~N.~C.}\
  \bibnamefont {Pfeifer}},\ }\emph {\bibinfo {title} {Simulation of Anyons
  Using Symmetric Tensor Network Algorithms}},\ \href@noop {} {Ph.D. thesis},\
  \bibinfo  {school} {The University of Queensland} (\bibinfo {year} {2011}),\
  \Eprint {http://arxiv.org/abs/1202.1522v2}
  {arXiv:1202.1522v2 [cond-mat.str-el]} \BibitemShut {NoStop}%
\bibitem [{Note1()}]{Note1}%
  \BibitemOpen
  \bibinfo {note} {Note that when two tensors are contracted together, all
  indices connecting these tensors should be listed consecutively. Sequences
  which do not satisfy this condition (i)~are always of suboptimal efficiency
  (see also \protect {Sec.~\ref {sec:suboptimal}}), and (ii)~will result in a
  warning being generated.}\BibitemShut {Stop}%
\bibitem [{\citenamefont {Evenbly}\ and\ \citenamefont
  {Pfeifer}()}]{evenbly2013}%
  \BibitemOpen
  \bibfield  {author} {\bibinfo {author} {\bibfnamefont {G.}~\bibnamefont
  {Evenbly}}\ and\ \bibinfo {author} {\bibfnamefont {R.~N.~C.}\ \bibnamefont
  {Pfeifer}},\ }\href@noop {} {}\Eprint {http://arxiv.org/abs/1310.8023} {arXiv:1310.8023 [cond-mat.str-el] (2013)}
  \BibitemShut {NoStop}%
\bibitem [{Note2()}]{Note2}%
  \BibitemOpen
  \bibinfo {note} {Note that \protect \textsc {Matlab}{} leaves trailing
  indices of dimension~1 implicit, and so will report the size of tensor~C as
  $3\times 3$ if queried, and not $3\times 3\times 1$. Equally, it is also
  valid to create tensor~C using the command \protect \texttt {C~=~rand(3,3);}
  instead of \protect \texttt {C~=~rand(3,3,1);} as given above.}\BibitemShut
  {Stop}%
\bibitem [{\citenamefont {Lam}\ \emph {et~al.}(1997)\citenamefont {Lam},
  \citenamefont {Sadayappan},\ and\ \citenamefont {Wenger}}]{lam1997}%
  \BibitemOpen
  \bibfield  {author} {\bibinfo {author} {\bibfnamefont {C.-C.}\ \bibnamefont
  {Lam}}, \bibinfo {author} {\bibfnamefont {P.}~\bibnamefont {Sadayappan}}, \
  and\ \bibinfo {author} {\bibfnamefont {R.}~\bibnamefont {Wenger}},\ }\href
  {\doibase 10.1142/S0129626497000176} {\bibfield  {journal} {\bibinfo
  {journal} {Parallel Processing Letters}\ }\textbf {\bibinfo {volume} {07}},\
  \bibinfo {pages} {157} (\bibinfo {year} {1997})}\BibitemShut {NoStop}%
\end{thebibliography}%

\end{document}